\documentclass[10pt]{article}

\newcommand{\hi}{{\cal{H}}}
\newcommand{\ma}{{\cal{M}}}
\newcommand{\dom}{{\cal{D}}}

\usepackage{conf_iap}

\begin{document}
\heading{%
%
Aspects of Information Theory in Curved Space\footnote{Talk presented at
the 10th Conference on General Relativity and Relativistic Astrophysics,
University of Guelph, Guelph, Ontario May 28-31, 2003.}\footnote{This work
was partially supported by the National Science and Engineering Research
Council of Canada.}

}
\par\medskip\noindent
\author{%
Achim Kempf$^1$
}
\address{%
Department of Applied Mathematics, University of Waterloo\\
Waterloo, Ontario N2L 3G1, Canada\\ akempf@uwaterloo.ca
}

\begin{abstract}
The often-asked question whether space-time is discrete or
continuous may not be the right question to ask: Mathematically,
it is possible that space-time possesses the differentiability
properties of manifolds as well as the ultraviolet finiteness
properties of lattices. Namely, physical fields in space-time
could possess a finite density of degrees of freedom in the
following sense: if a field's amplitudes are given on a
sufficiently dense set of discrete points then the field's
amplitudes at all other points of the manifold are fully
determined and calculable. Which lattice of sampling points is
chosen should not matter, as long as the lattices' spacings are
tight enough, for example, not exceeding the Planck distance. This
type of mathematical structure is known within information theory,
as sampling theory, and it plays a central role in all of digital
signal processing.
\end{abstract}
\section{Introduction}
Let us reconsider why the unification of general relativity and
quantum theory has proven so difficult. Mathematically, the
problems clearly begin with the fact that the two theories are
formulated in the quite different languages of differential
geometry and functional analysis. Physically, an important problem
appears to be that general relativity and quantum theory, when
considered together, are indicating that the notion of distance
loses operational meaning at the Planck scale of about $10^{-35}m$
(assuming 3+1 dimensions). Namely, if one tries to resolve a
spatial structure with an uncertainty of less than a Planck
length, then the corresponding momentum uncertainty should
randomly curve and thereby significantly disturb the very region
in space that is meant to be resolved.

One of the problems in the effort of finding a unifying theory of
quantum gravity is, therefore, to develop a mathematical framework
which combines differential geometry and functional analysis such
as to give a precise description of a notion of a shortest
distance in nature. Candidate theories may become testable when
introduced to inflationary cosmology and compared to the CMB
measurements, see \cite{cosm}.

In the literature, there has been much debate about whether the unifying
theory will describe space-time as being discrete or continuous. It is
tempting, also, to speculate that a quantum gravity theory such as M
theory, see e.g. \cite{pol}, a noncommutative geometric theory, see e.g.
\cite{connesmajid}, or a foam theory, see e.g. \cite{foam}, once fully
understood, might reveal the structure of space-time as being in some
sense in between discrete and continuous, possibly such as to combine the
the differentiability of manifolds with the ultraviolet finiteness of
lattices. At first sight, this third possibility seems to be ruled out,
however: as G{\"o}del and Cohen proved, no set can be explicitly
constructed whose cardinality would be in between discrete and continuous,
see e.g. \cite{ak-paris}.

The message of this talk is that, nevertheless, there still is at least
one mathematical possibility by which a theory of quantum gravity might
yield a description of space-time which combines the differentiability of
manifolds with the ultraviolet finiteness of lattices:

\section{Fields with a finite density of degrees of freedom}

Let us recall that physical theories are formulated not directly
in terms of points in space or in space-time but rather in terms
of the functions in space or in space-time. This suggests a whole
new class of mathematical models for a finite minimum length.

Namely, fields in space-time could be functions over a
differentiable manifold as usual, while, crucially, the class of
physical fields is such that if a field is sampled only at
discrete points then its amplitudes can already be reconstructed
at \it all \rm points in the manifold - if the sampling points are
spaced densely enough. The maximum average sample spacing which
allows one to reconstruct the continuous field from discrete
samples could be on the order of the Planck scale, see
\cite{ak-prl}.

Since any one of all sufficiently tightly spaced lattices would
allow reconstruction, no particular lattice would be preferred. It
is because no particular lattice is singled out that the symmetry
properties of the manifold can be preserved.

The physical theory, i.e. fields and actions etc. could be written,
equivalently, either as living on a differentiable manifold, thereby
displaying e.g. external symmetries, or as living on any one of the
sampling lattices of sufficiently small average spacing, thereby
displaying its ultraviolet finiteness. Physical fields, while being
continuous or even differentiable, would possess only a finite density of
degrees of freedom.

The mathematics of classes of functions which can be reconstructed
from discrete samples is well-known, namely as \it sampling
theory, \rm in the information theory community, where it plays a
central role in the theory of sources and channels of continuous
information as developed by Shannon, see \cite{shannon}.
\section{Sampling theory}
The simplest example in sampling theory is the Shannon sampling
theorem: Choose a frequency $\omega_{max}$. Consider the class
$B_{\omega_{max}}$ of continuous functions $f$ whose frequency
content is limited to the interval $(-\omega_{max},\omega_{max})$,
i.e. for which:
$ \tilde{f}(\omega)~=~\int_{-\infty}^{\infty} f(x) e^{-i\omega x}
= 0 \mbox{~~~whenever ~~} \vert\omega\vert\ge\omega_{max}.$
If the amplitudes $f(x_n)$ of such a function are known at
equidistantly spaced discrete values $\{x_n\}$ whose spacing is
$\pi/\omega_{max}$ or smaller, then the function's  amplitudes
$f(x)$ can be reconstructed for all $x$. The reconstruction
formula is:
\begin{equation}
f(x)~ =~ \sum_{n=-\infty}^\infty ~f(x_n) ~
\frac{\sin[(x-x_n)\omega_{max}]}{(x-x_n)\omega_{max}}
\end{equation}
The theorem is in ubiquitous use in digital audio and video as
well as in scientific data taking. Sampling theory, see
\cite{ferreira}, studies generalizations of the theorem for
various different classes of functions, for non-equidistant
sampling, for multi-variable functions and it investigates the
effect of noise, which could be quantum fluctuations in our case.
As was shown in \cite{ak-prl}, generalized sampling theorems
automatically arise from stringy uncertainty relations, namely
whenever there is a finite minimum position uncertainty $\Delta
x_{min}$, as e.g. in uncertainty relations of the type: $\Delta x
\Delta p \ge \frac{\hbar}{2}(1 + \beta (\Delta p)^2 +...)$, see
\cite{ucr}. A few technical remarks: the underlying mathematics is
that of symmetric non self-adjoint operators. Through a theorem of
Naimark, unsharp variables of POVM type arise as special cases.

\section{Information Theory on Curved Space}
Let us consider as a natural (because covariant) analogue of the bandwidth
restriction of the Shannon sampling theorem in curved space the presence
of a cutoff on the spectrum of the Laplace operator $-\Delta$ on a
Riemannian manifold $\ma$ (or the d'Alembert or the Dirac operator on a
pseudo-Riemannian or a spin manifold respectively).

We start with the usual Hilbert space $\hi$ of square integrable
scalar functions over the manifold, and we consider the dense
domain $\dom\subset\hi$ on which the Laplacian is essentially
self-adjoint. Using physicists' sloppy but convenient terminology
we will speak of all points of the spectrum as eigenvalues,
$\lambda$, with corresponding ``eigenvectors" $\vert \lambda)$.
Since we are mostly interested in the case of noncompact
manifolds, whose spectrum will not be discrete, some more care
will be needed, of course. For Hilbert space vectors we use the
notation $\vert \psi)$, in analogy to Dirac's bra-ket notation,
only with round brackets.

Let us define $P$ as the projector onto the subspace spanned by the
eigen\-spaces of the Laplacian with eigenvalues smaller than some fixed
maximum value $\lambda_{max}$. (For the d'Alembertian and for the Dirac
operator, let $\lambda_{max}$ bound the absolute values of the
eigenvalues.)

We consider now the possibility that in nature all physical fields are
contained within the subspace $\dom_s= P.\dom$, where $\lambda_{max}$
might be on the order of $1/l^2_{Planck}$. In fact, through this spectral
cutoff, each function in $\dom_s$ acquires the sampling property: if its
amplitude is known on a sufficiently dense set of points of the manifold,
then it can be reconstructed everywhere. Thus, through such a spectral
cutoff a sampling theorem for physical fields arises naturally. To see
this, assume for simplicity that one chart covers the $N$-dimensional
manifold. Consider the coordinates $\hat{x}_j$, for $j=1,...,N$ as
operators that map scalar functions to scalar functions: $\hat{x_j}:
\phi(x) \rightarrow x_j \phi(x)$. On their domain within the original
Hilbert space $\hi$, these operators are essentially self-adjoint, with an
``Hilbert basis" of non-normalizable joint eigenvectors $\{\vert x)\}$. We
can write scalar functions as $\phi(x)=( x\vert \phi)$, i.e. scalar
functions are the coefficients of the abstract Hilbert space vector $\vert
\phi)\in \hi$ in the basis of the vectors $\{\vert x )\}$. The continuum
normalization of the $\vert x)$ is with respect to the measure provided by
the metric. On the domain of physical fields, $\dom_s$, the multiplication
operators $\hat{x}_j$ are merely symmetric but not self-adjoint. The
projections $P\vert x)$ of the eigenvectors $\vert x )$ onto the physical
subspace $\dom_s$ are in general no longer orthogonal. Correspondingly,
the uncertainty relations are modified, see \cite{ak-old}.

Consider now a physical field, i.e. a vector $\vert \phi)\in
\dom_s$, which reads as a function: $\phi(x)=(x\vert\phi)$. Assume
that only at the discrete points $\{x_n\}$ the field's amplitudes
$\phi(x_n)=\langle x_n\vert \phi\rangle$ are known. Then, if the
discrete sampling points $\{x_n\}$ are sufficiently dense, they
fully determine the Hilbert space vector $\vert \phi\rangle$, and
therefore $\phi(x)$ everywhere. To be precise, we assume the
amplitudes
\begin{equation}
\phi(x_n)=(x_n\vert\phi)=\sum_{\vert\lambda\vert<\lambda_{max}}
\!\!\!\!\!\!\!\!\!\!\!\!\!\!\int ~(x_n\vert \lambda)(\lambda \vert
\phi)~d\lambda
\end{equation}
 to be known. We use the sum and
integral notation because $\{\lambda\}$ may be discrete and or continuous
(the manifold $\ma$ may or may not be compact). Define $K_{n\lambda}=
(x_n\vert\lambda)$. The set of sampling points $\{x_n\}$ is dense enough
for reconstruction iff $K$ is invertible, because then: $(\lambda\vert
\phi)=\sum_n K^{-1}_{\lambda,n} \phi(x_n)$ and we therefore obtain the
reconstruction formula:
\begin{equation}
\phi(x) = \sum_n \left(\sum_{\vert\lambda\vert<\lambda_{max}}
\!\!\!\!\!\!\!\!\!\!\!\!\!\!\int ~ (x\vert\lambda) K^{-1}_{\lambda
n}~d\lambda\right) ~\phi(x_n)
\end{equation}
In communication theory, the stability of the reconstruction is important
due to noise and is handled as in \cite{landau}. Here, not only may
quantum fluctuations act as `noise', but information can also be
entangled. Still, following Shannon and Landau, it is natural to define
the density of degrees of freedom through the number of dimensions of the
space of functions in $\dom_s$ with essential support in a given volume.
Clearly, we recover conventional Shannon sampling as a special case. The
Shannon case has been applied to inflationary cosmology in \cite{cosm} for
flat space. It should be very interesting to apply to cosmology also the
general approach presented here, both to generic non-flat spatial slices,
and also to the fully covariant case based on a cutoff of the spectrum of
the Dirac or d'Alembert operator. In particular, the analysis of the
analog of sampling theory in the case of indefinite metrics should provide
a new approach to the problem of generally covariant UV cutoffs.

We also note that higher than second powers of the fields (second powers
occur as scalar products in the Hilbert space of fields) are now
nontrivial in quantum field theoretical actions: This is because the
multiple product of fields needs to be defined such as to yield a result
within the cut-off Hilbert space. In this context, it should be
interesting also to reconsider the mechanism of Sakharov's induced
gravity, see \cite{sakharov}.

A sampling theoretical cutoff can be applied in arbitrary dimensions and
it should of interesting, e.g., to model a maximum achievable information
density on black hole horizons this way. For holography, see e.g.
\cite{bekensteinetc}. Note that the sampling theoretical cutoff is always
holographic in the sense that all information is encoded already in
zero-dimensional sets, namely in any set of sampling points from which
reconstruction is possible. In principle, holography in this sense should
not be surprising: Any quantum theory which lives on a separable Hilbert
space, in any space-time dimension, lives on a Hilbert space with a
countable basis. For example, the Hilbert space of ordinary QM in three
dimensions is unitarily equivalent to the Hilbert space of QM in any other
number of dimensions, simply because all separable Hilbert spaces are
unitarily isomorphic. The key observation in our sampling theory approach
is that discrete sets of formal position eigenvectors can be chosen as
such a countable basis in the Hilbert space, if there is bandwidth cutoff.

Our approach to sampling on curved space significantly simplifies in the
case of compact manifolds, where the spectrum of the Laplacian is discrete
and the cut off Hilbert space $\hi_s$ is finite dimensional. Intuitively,
it is clear that knowledge of a function at as many points as is the
dimension of the cutoff Hilbert space generically allows one to
reconstruct the function everywhere. If the compact manifold is a group,
the Peter-Weyl theorem provides us explicitly with the finite-dimensional
Hilbert spaces of functions of ``finite bandwidth" obtained by cutting of
the spectra of Casimir operators. In the particular case of $SU(2)$ and
the Laplacian we obtain the fuzzy sphere, see e.g. \cite{madore}, which
has been much discussed in the context of noncommutative geometry,
\cite{connesmajid}.

In the literature, sampling theory on generic Riemannian manifolds has
been little studied so far. This is because sampling theory originated and
finds most of its applications in communication engineering. Interesting
results that are of relevance here were obtained, however, by Pesenson,
see e.g. \cite{pesenson}, who considered, in particular, the case of
homogeneous manifolds. In \cite{pesenson}, the starting point is also a
cutoff on the Laplace operator's spectrum. Reconstruction, however, works
differently, namely by approaching the solution iteratively in a Sobolev
space setting.

Much about the Laplace operator has been established in the field of
spectral geometry, see e.g. \cite{specs1,specs2,specs3,specs4} and these
methods and results should be very useful for our approach described here.

\vfill
\end{document}